\documentstyle[floats,twocolumn,aps,psfig,prl]{revtex}

\begin{document}
\draft

\twocolumn[\hsize\textwidth\columnwidth\hsize\csname @twocolumnfalse\endcsname


\title{Superconducting energy gap distribution of MgB$_2$ investigated by
point-contact spectroscopy}

\author{F. Laube$^1$, G. Goll$^1$, J. Hagel$^ 1$, H. v. L\"ohneysen$^{1,2}$,
D. Ernst$^2$, T. Wolf$^2$}

\address{$^1$Physikalisches Institut, Universit\"at Karlsruhe,
D-76128 Karlsruhe, Germany\\
$^2$ Forschungszentrum Karlsruhe, Institut f\"ur Festk\"orperphysik,
D-76021 Karlsruhe, Germany}

\date{\today}
\maketitle

\begin{abstract}
We performed point-contact spectroscopy on the binary superconductor
MgB$_2$. The differential conductance shows gap-related structures which
vary in width and position from contact to contact. The distribution of energy 
gaps shows a distinct
accumulation around 1.7 and 7\,meV which is associated with the occurrence of
a small and a large energy gap in MgB$_2$. While with increasing $T$
the structure in $dI/dV$ associated with the small gap  is present up to $T_C$,
in magnetic field it is suppressed well below $B_{c2}$.
\end{abstract}

\pacs{PACS numbers: 74.70.Ad, 74.80.Fp, 74.20.Rp}
\vskip2pc]

The observation of superconductivity at 39\,K in the simple binary compound
MgB$_2$ \cite{Nag01} was a surprise for most condensed matter scientists
and immediately raises the question of the mechanism of superconductivity
in this material. Following a basic approach to superconductivity, 
two key issues have to be considered: the order-parameter symmetry
and the coupling mechanism. For the latter, a significant boron isotope effect
was observed in MgB$_2$ \cite{Bud01} which is consistent with a phonon-mediated
BCS mechanism where the boron phonon modes are playing an important role.
A moderatly strong electron-phonon coupling constant $\lambda_{\rm 
el-ph}\approx 0.8$ was derived
from measurements of the specific heat \cite{Kre01} in good agreement with
recent theoretical predictions \cite{Liu01,Kon01,Boh01}.

The order-parameter symmetry has already been investigated by tunneling
spectroscopy and point-contact spectroscopy (PCS) 
\cite{Bug01,Rub01,Kar01,Sha01,Sch01,Che01,Ple01,Giu01,Sza01}. Although the 
spectra show unambiguous features
of an energy gap in the density of states, probably with $s$-wave symmetry of 
the
order parameter, the results are controversal on the gap width. Values of
$2\Delta /k_{\rm B}T_c$ ranging from 1.2 to 4.7 have been reported, raising the
possibility of an anisotropic energy gap or even multiple gaps. Careful 
analysis of
specific-heat measurements \cite{Yan01,Bou01} and photoemission spectroscopy
\cite{Tsu01} support this scenario as well.

\begin{figure}[ht]
  \centerline{\psfig{file=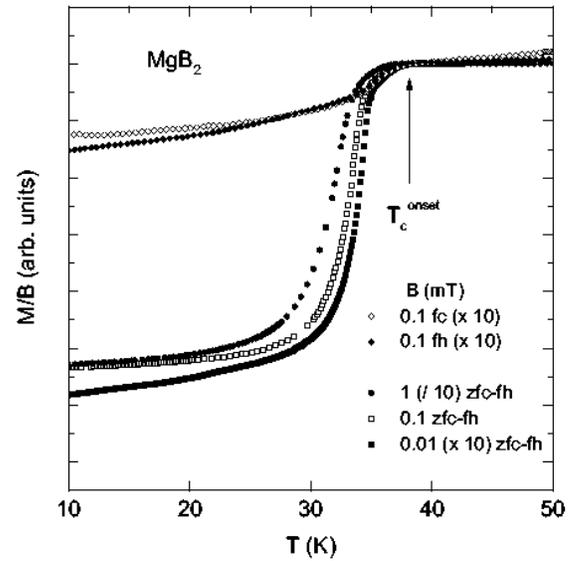,clip=,width=7.5cm}}
  \caption[DC]{DC magnetization of MgB$_2$ in different applied magnetic 
fields.}
\label{DC}
\end{figure}

We present measurements of the differential conductance $dI/dV$ of normal 
conductor/superconductor point contacts between a Pt 
tip and a MgB$_2$ pellet. 
The MgB$_2$ pellet was prepared in an Ar atmosphere from magnesium 
filings 
($>99.9$\,\%, Chempure) and powdered boron (99\,\%, Ventron). The powders were 
mixed 
in the ratio Mg:B = 1.1:2 and pressed into a pellet.  After wrapping the pellet
in a Ta foil it was enclosed in an Fe cylinder and inserted in a high-pressure 
furnace. The pellet was then heated to 1073\,K for 1.5\,h and to 953\,K for 2h 
under an Ar pressure of 58\,MPa. The DC magnetization of the pellet 
(Fig.\,\ref{DC}) indicates an onset superconducting transition temperature 
$T_c^ {onset}=38.2$\,K. Almost the full theoretical shielding signal is reached
in 1\,mT applied field in a zero-field cooled (zfc) -- field heated (fh) temperature 
cycle. The Meissner signal remains weak, i.\,e. of the order of a few percent.
Upon reducing
the applied field to $10\,\mu$T, a kink in the transition curve at 35\,K
reveals the granular structure of the sample: below $T_c^ {onset}$ first the
individual grains become superconducting, below 35\,K intergranular shielding 
currents cause shielding of the whole pellet.

\begin{figure}[ht]
  \centerline{\psfig{file=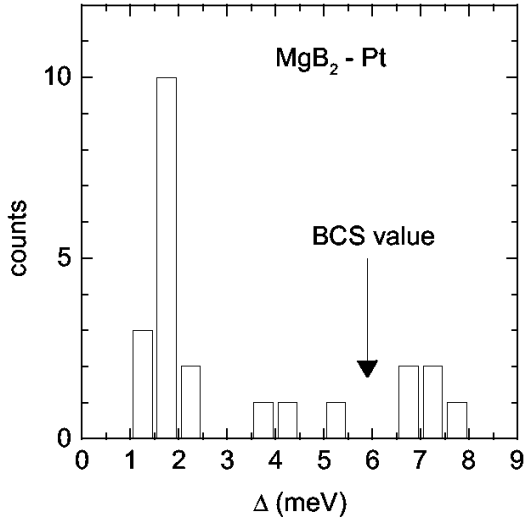,clip=,width=7.5cm}}  
\caption[hist]{Extracted values of the gap energy for different MgB$_2$ 
contacts}
\label{hist}
\end{figure}

\begin{figure}[t]
 \centerline{\psfig{file=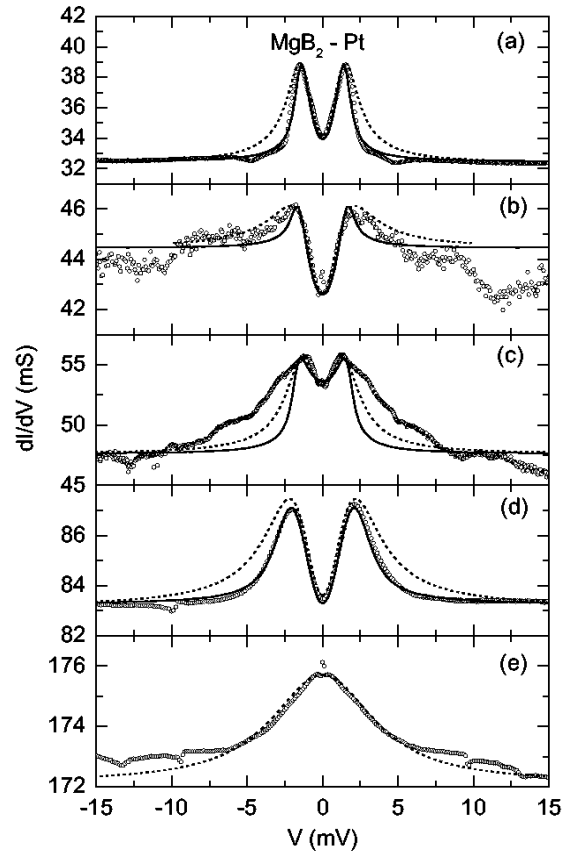,clip=,width=8cm}}
\caption[mitohne]{Measured $dI/dV$ curves and 
BTK fits to the small gap structures for different MgB$_2$/Pt 
contacts at $T=1.7$\,K (panel a-c) and 5\,K (panel d and e) either
with thermal smearing only ($\Gamma = 0$, solid lines)
or with additional finite-lifetime smearing ($\Gamma\neq 0$, dashed lines).
The fit parameters are summarized in Table\,{\protect\ref{para}}.}
\label{mitohne}
\end{figure}

PCS spectra were taken on as grown and polished surfaces of the MgB$_2$ pellet.
Different contacts were established 
in situ at low temperatures in liquid Helium. The PCS spectra, i.\,e. 
the differential resistance $dV/dI$ vs
$V$ were obtained using a standard 
four-point lock-in technique. The zero-bias resistance $R_0$ of stable contacts
ranges from  1 to 60 $\Omega$. 
$dI/dV$ vs $V$ curves are obtained by numerical inversion of the $dV/dI$ data.
The $dI/dV$ curves taken at low $T\ll T_c$ (between 1.7 and 5\,K, cf.
Fig.\,\ref{mitohne}), 
show most often maxima at voltage $\pm V_0$, presumably related to the superconducting
energy gap $\Delta$.  
The position
of the maxima $\Delta=eV_0$, varies depending on the location of the Pt needle
on the polycrystalline sample
where the point contact was established. The histogram depicted in 
Fig.\,\ref{hist} summarizes the position of maxima of 20 contacts.
In a few cases a single maximum
centered at $V=0$ was observed (Fig.\,\ref{mitohne}e).
For these cases the value of
$\Delta_0=\Delta(T\rightarrow 0)$ was derived from the modified BTK fit (see below) and is also
included in Fig.\,\ref{hist}.
This histogram in nice consistency reflects the gap distribution
reported in literature.
One dominant feature is a maximum at about 1.7\,meV, a second occurs at 
about 7\,meV. 

Following the BTK theory \cite{Blo82} of transport
through a metallic microcontact between a normal metal (N) and an isotropic
superconductor (S), Andreev reflection is the dominant transport process
at the N-S interface. Andreev-reflected charge carriers lead to an excess
current which increases the point-contact conductance. Depending
on the strength of the interface barrier a characteristic double-maximum
feature is expected in $dI/dV$, with the maxima at $\pm \Delta /e$ where
$\Delta$ is the energy gap. Inelastic scattering in the contact 
region usually leads to a smearing and weakening of the gap-related features.
This can be empirically taken into account by replacing the quasiparticle
energy $E$ with $E+i\Gamma$, where $\Gamma$ parametrizes
the finite lifetime of the quasiparticles \cite{Dyn78}.
We observed a few contacts where the $dI/dV$ curves show maxima at $\pm V_0$ and
$\pm V_1$. These curves can be regarded as a superposition of two
double-maximum features. Consequently, the maxima are interpreted as
being associated with two different 
energy gaps and the presence of multigap superconductivity in
MgB$_2$.

Multigap superconductivity in MgB$_2$ has recently theoretically 
been suggested by Liu {\it et al.} \cite{Liu01} based on the electronic
structure proposed by Kortus {\it et al.} \cite{Kor01}:
The Fermi surface consists of two sets of sheets, two-dimensional (2D)
cylindrical sheets and three-dimensional (3D) tubular network. In the clean
limit, two different superconducting order parameter are predicted: a larger
one on the 2D part of the Fermi surface, a smaller one (about 1/3) on the 3D
part of the Fermi surface. The multi-sheetedness of the Fermi surface
should lead to a substantial difference between the in-plane 
and out-of-plane tunneling spectra. The effect of strong defect scattering
drives the system into the dirty limit where the two energy gaps are mixed
and an isotropic BCS gap is obtained. Quite remarkable is the effect on the
temperature dependence of the gap in both limits. In the clean limit both
energy gaps close at $T_c$ while in the dirty limit the resultant isotropic
gap is substantially suppressed.

\begin{table}
\begin{tabular}{cccccccc}
   &   &\multicolumn{3}{c}{solid lines}&\multicolumn{3}{c}{dashed lines}\\
   &$T$ (K)&$\Delta_0$ (meV)&$Z$&$r$&$\Delta_0$ (meV)&$\Gamma$ (meV)&$Z$\\
(a)& 1.7 & 1.5 & 0.5  & 0.368 & 1.5 & 0.65 & 0.57 \\
(b)& 1.7 & 1.7 & 0.8  & 0.079 & 1.1 & 1.4  & 0.9  \\
(c)& 1.7 & 1.7 & 0.3  & 0.305 & 1.5 & 0.87 & 0.47 \\
(d)& 5   & 2.1 & 0.615& 0.128 & 1.3 & 1.3  & 0.73 \\
(e)& 5   & --  &  --  &   --  & 1.1 & 3.6  & 0.1  \\
\end{tabular}
\caption{Fit parameter of the theoretical curves in Fig.\,{\protect\ref{mitohne}}}
\label{para}
\end{table}
In the following we focus on spectra which show features associated with
the small gap structure only. Fig.\,\ref{mitohne}a-e show a number of spectra
of this type together with theoretical curves calculated within the
modified BTK theory either with thermal smearing only ($\Gamma = 0$, solid lines)
or with additional finite-lifetime smearing ($\Gamma\neq 0$, dashed lines). 
The fit parameters of these curves are summarized in Table\,\ref{para}.
Although the width of the spectra is almost the same, the
voltage dependence is different. Curve a and d (referred to as type 1) 
are described
very well with a pure ($\Gamma=0$) BTK fit with an energy gap (assumed to be
isotropic)
$\Delta_0=1.5$\,meV (a) and $\Delta_0=2.1$\,meV (d), a 
barrier strength $Z=0.5$ (a) and $Z=0.615$ (d), respectively.
The fits have been scaled by a factor of 
$r=[dI/dV(0)-dI/dV(15\,{\rm mV})]_{\rm exp}/[dI/dV(0)-dI/dV(15\,{\rm mV})]_{\rm
theo}=0.368$ (a) and 0.128(d), respectively, 
in order to meet the absolute magnitude of the experimental conductance change
$[dI/dV(0)-dI/dV(15\,{\rm mV})]_{\rm exp}$.
The common feature of curve b, c, and e (referred to as type 2) is that the 
decrease of the 
conductance for $V\geq\Delta/e$ is weaker than expected from a pure BTK fit,
i.\,e. the experimental curves are much wider. Smearing due to inelastic
scattering in the contact region can be one origin
of wider structures which has been modelled by $\Gamma\neq 0$.
The parameters $\Delta$,
$\Gamma$, and $Z$ are chosen in such a way that the calculated curves describe
the experimental conductance change without the necessity of scaling.
Introducing of $r$ as an additional fit parameter to improve the fit would
require an unphysical value $r>1$.
A relatively large $\Gamma$ of the same order of 
magnitude as $\Delta$ or even larger is needed to model the curves, and the
resulting description of the spectra is not really satisfactory.
The failure of the 
model in describing curve c hints at a second, intrinsic origin, namely an
apparent gap
distribution due to an anisotropic small gap. 
This means that in addition to the occurrence of two energy gaps in MgB$_2$,
these gaps themselves might not be isotropic. Recently
Giubileo {\it et al.} \cite{Giu01} pointed out that the gap width varies even in
the case where spatial inhomogeneities can be ruled out as an origin of a gap
distribution. It might be of interest to note that all three curves show a weak conductance maximum around
zero-bias which might be caused by nodes of the gap function \cite{Che01}.

\begin{figure}[t]
 \centerline{\psfig{file=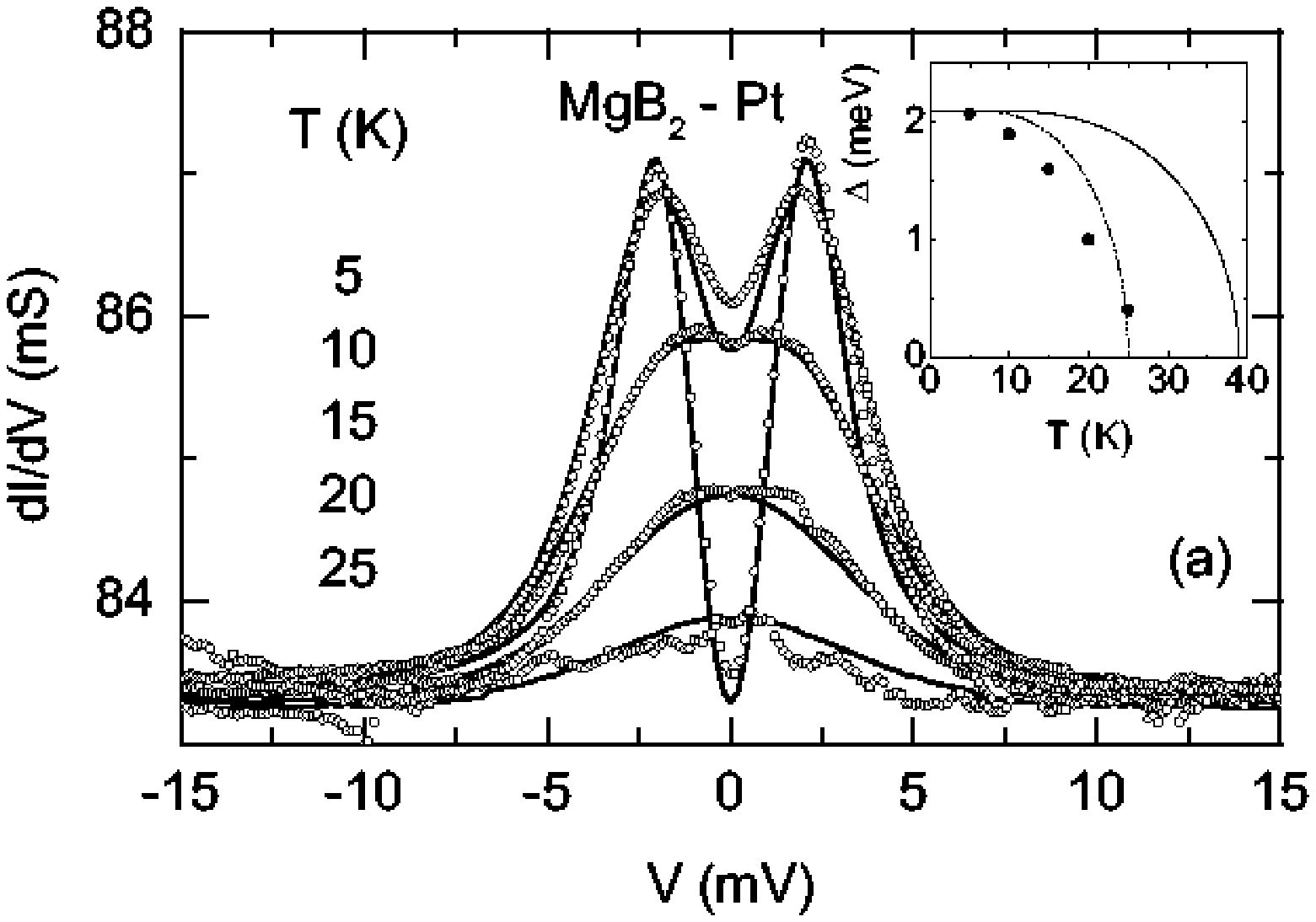,clip=,width=8cm}}
  \centerline{\psfig{file=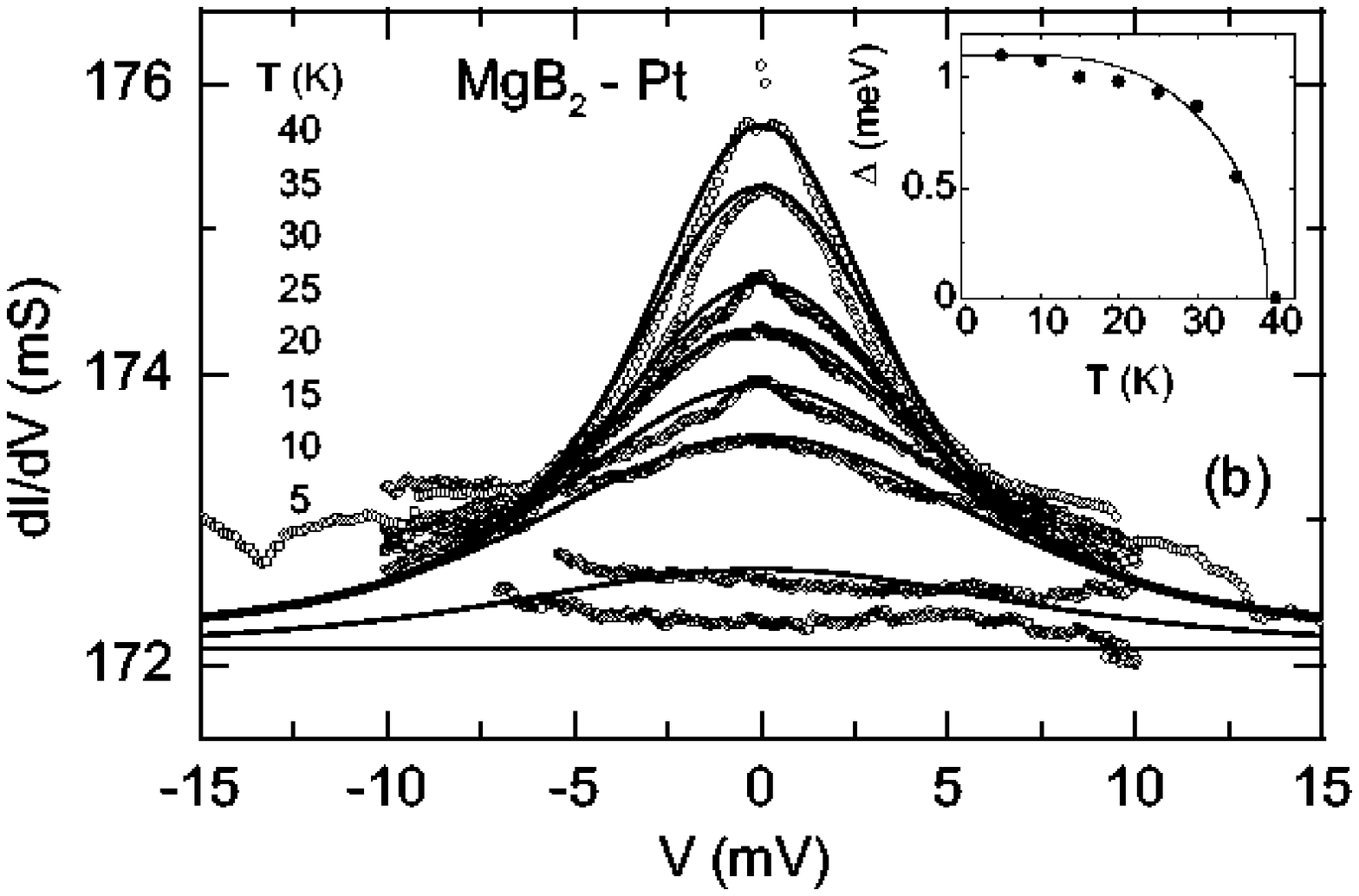,clip=,width=8cm}}
\caption[temp]{Temperature dependence of point contacts of type 1 (a) and 
2 (b).}
\label{temp}
\end{figure}

In order to further elucidate the difference of the spectra we now turn to the 
temperature dependence of both types depicted in Fig.\,\ref{temp} a and b
together with theoretical curves. With increasing temperature the gap-related 
structure weakens in both cases. However, while gap-related features of the 
spectra of type 2 vanish close to the bulk $T_c$ and almost follow 
the BCS weak-coupling behavior of $\Delta (T)$, the isotropic gap extracted from spectra
of type 1 closes at $\approx 0.7\,T_c$. Without overrating the experimental
facts, the results resemble the proposed multigap scenario. Bearing in mind the
granular structure of the sample and the unknown microscopic structure of the point
contact, it is conceivable that both clean and dirty limits are locally
realized on different parts of the sample.

\begin{figure}[t]
  \centerline{\psfig{file=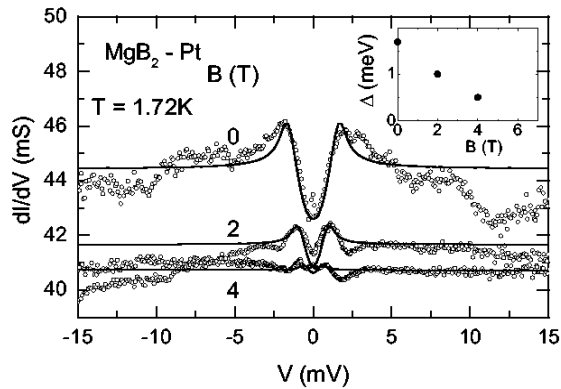,clip=,width=8cm}}  
\caption[feld]{Magnetic field dependence of the point contact shown in
Fig.\,{\protect\ref{mitohne}}b. The inset shows
$\Delta(B)$ extracted from the BTK fit to the data.}
\label{feld}
\end{figure}

In an applied magnetic field $B$ the gap structure weakens
with increasing field and vanishes around 6\,T (see Fig.\,\ref{feld}). 
The small gap is much more sensitive to an applied field than expected from
estimates of the upper critical field $B_{c2}(0)$ from
resistivity \cite{Fin01} and magnetization 
measurements \cite{Lar01}. These data suggest that a field well above 12\,T is
necessary to suppress superconductivity at 1.7\,K.
Szabo {\it et al.} have recently reported an even stronger field dependence
of the small gap \cite{Sza01}. In line with specific-heat measurments
\cite{Yan01,Bou01}, $B=1$\,T substantially suppresses the small gap.
The result hints at a different nature of the two energy gaps probably due to
their different dimensionality.

In conclusion, we have examined point contacts on MgB$_2$ in order to 
investigate the superconducting energy gap. The gap width obtained from
$dI/dV$ vs $V$ spectra varies from contact to contact depending on the
microscopic structure and grain orientation in the point-contact region.
The result is compatible with multigap superconductivity with a small gap
present up to $T_c$.


\end{document}